\documentclass[journal,10pt]{IEEEtran}

\usepackage{color}  
\usepackage[final]{graphicx} 
\usepackage{amsmath,amssymb,amscd,latexsym,dsfont,amsthm,algorithm,algorithmic,amsbsy,epsfig,bbm,mathrsfs,fancyhdr,fancyvrb} 
\usepackage{bmpsize}
\usepackage{upgreek,textgreek}
\usepackage{subcaption}
\usepackage{caption}
\usepackage{psfrag}
\usepackage{epstopdf} 
\usepackage{tabls}
\usepackage[noadjust]{cite}
\usepackage{color,comment}
\usepackage{balance}
\usepackage{float}
\usepackage{placeins}
\usepackage{lipsum}
\usepackage[top=0.75in, bottom=1in, left=0.625in, right=0.625in]{geometry}
\usepackage{algorithm} 
\usepackage{algorithmic}  
\usepackage[algo2e]{algorithm2e} 

\newcommand{\tr}[1]{\mathrm{#1}}

\newcounter{tempEquationCounter}
\newcounter{thisEquationNumber}
{\setcounter{equation}{\value{tempEquationCounter}}
\end{figure*}
}%

\begin{document}
\title{Interference Management with Partial Uplink/Downlink Spectrum Overlap}

\author{%
Itsikiantsoa Randrianantenaina, Hesham Elsawy, Hayssam Dahrouj, and ~Mohamed-Slim~Alouini\\
\thanks{%
I. Randrianantenaina, H. Elsawy and M.-S. Alouini are with the Computer, Electrical, and Mathematical Science and Engineering (CEMSE) Division, King Abdullah University of Science and Technology (KAUST), Thuwal, Saudi Arabia. H. Dahrouj is with the department of Electrical and Computer Engineering at Effat University, Jeddah, Saudi Arabia. [e-mails: \{itsikiantsoa.randrianantenaina, hesham.elsawy, slim.alouini\}@kaust.edu.sa], hayssam.dahrouj@gmail.com}%
\thanks{%
This work was funded in part by King Abdullah University of Science and Technology (KAUST). Hayssam Dahrouj would like to thank Effat University in Jeddah, Saudi Arabia, for funding the research reported in this paper through the Research and Consultancy Institute.
}%
}


\maketitle
\thispagestyle{empty}

\begin{abstract}

Simultaneous reuse of spectral resources by uplink and
downlink, denoted as in-band full duplex (FD) communication, is
promoted to double the spectral efficiency when compared to its
half-duplex (HD) counterpart. Interference management, however,
remains challenging in FD cellular networks, especially when
high disparity between uplink and downlink transmission powers
exists. The uplink performance can be particularly deteriorated
when operating on channels that are simultaneously occupied
with downlink transmission. This paper considers a cellular
wireless system with partial spectrum overlap between the
downlink and uplink. The performance of the system becomes,
therefore, a function of the overlap fraction, as well as the power
level of both the uplink and downlink transmissions. The paper
considers the problem of maximizing an overall network utility to
find the uplink/downlink transmission powers and the spectrum
overlap fraction between the uplink and downlink spectrum in
each cell, and proposes solving the problem using interior point
method. Simulations results confirm the vulnerability of the uplink
performance to the FD operation, and show the superiority of
the proposed scheme over the FD and HD schemes. The results
further show that explicit uplink and downlink performance
should be considered for efficient design of cellular networks with
overlapping uplink/downlink resources.

\end{abstract}

\section{Introduction}

Recent advancement in design of transceivers with self-interference cancellation (SIC) capabilities triggers nowadays interests in the in-band full duplex (FD) communication. In particular, FD transceivers can simultaneously transmit and receive on the same channel by using SIC techniques to provide sufficient protection for the receive chain from the overwhelming transmit signal~\cite{Hong2014,Lee2015, Sabharwal2014}. When compared to the half-duplex (HD) communication which split the spectrum between forward and reverse links, FD communication offers higher spectrum utilization by simultaneously reusing the entire spectrum by both links. FD communication specifically combines disjoint forward link and reverse link HD channels to common FD channels, this operation doubles  the bandwidth (BW) available on each link. Depending on the SIC efficiency, FD communication is expected to provide up to $100\%$ capacity gains when compared to its HD counterpart~\cite{Hong2014, Does2015Xie}.

In the context of large-scale networks, SIC is not the only challenge for harvesting the foreseen FD communication gains. Due to spatial frequency reuse, FD communication experiences higher levels of interference on the common FD channels when compared to the disjoint HD channels, due to the induced forward-reverse interference. While HD receivers experience intra-mode interference only, FD receivers experience intra-mode, cross-mode (i.e., interference between forward and reverse links), and self interference. While SIC solves the self interference part, efficient interference mitigation techniques are required to solve the inter-mode/cross-mode interference parts.  Otherwise, cross-mode interference would diminish the higher BW gains offered by FD communication~\cite{Does2015Xie, Full2015Mohammadi, Analyzing2013Goyal}. The cross-mode interference problem is far more acute in cellular networks, especially when disparity between the transmission powers in the forward link (downlink) and the reverse link (uplink) exits ~\cite{Itsikiantsoa2015, Alammouri2015}. It is shown that FD can indeed increase the spectral efficiency for downlink transmissions~\cite{Alves2014, Lee2015, Alammouri2015 }. However, cross-mode interference can cause intolerable deterioration in the uplink performance~\cite{Itsikiantsoa2015, Alammouri2015}.  Therefore, efficient interference mitigation techniques are required to balance the tradeoff between uplink and downlink performance and, at the same time, maximize the overall network performance.

For efficient FD operation in cellular networks, efforts are invested to mitigate the cross-mode interference induced by FD communication \cite{Huberman2014,Nguyen2015,Full2015Goyal}. In \cite{Huberman2014}, precoding schemes maximizing the spectral and energy efficiencies of a  multiuser MIMO FD network are proposed. Power control algorithms  for FD communication are suggested in \cite{Nguyen2015} in order to overcome interferences. The authors in \cite{Full2015Goyal} manage interference by different types of scheduling to coordinate transmission in nearby cells. However, in these works, only fully overlapping uplink/downwlink channels are considered which might not be the optimal choice. Moreover, the authors do not analyze separately the uplink and downlik performance which  hide the damage uplink transmission might suffer of.

The authors in \cite{Alammouri2015} propose the $\alpha$-duplex scheme in which the cross mode interference is controlled via the parameter $\alpha$, which can be seen as the fraction of overlap between uplink/downlink spectrum. Reference \cite{Alammouri2015}, especially, shows that with the proper choice of pulse-shapes and parameter $\alpha$, appreciable simultaneous improvement in the UL and DL rates can be achieved. However, the system in \cite{Alammouri2015} use a fixed value of $\alpha$ for all BSs. Further, reference \cite{Alammouri2015} does not optimize the uplink and downlink power levels, which results in a non-optimized FD operation.

This paper considers a cellular wireless system with a partial spectrum overlap between the downlink and uplink. The performance of the system becomes, therefore, a function of the overlap fraction, as well as the power level of both the uplink and downlink transmissions. Unlike reference \cite{Alammouri2015} which considers the problem from a statistical  perspective, the main contribution of this paper is to maximize an overall network utility to find the uplink/downlink transmission powers and the spectrum overlap fraction between the uplink and downlink spectrum in each cell. The paper proposes solving the problem using interior point method. Our study confirms the vulnerability of uplink performance to the FD operation and shows the superiority of the $\alpha$ duplex scheme over the FD and HD schemes. It also shows the conservative figures reported by the statistical study in \cite{Alammouri2015} regarding the FD gains. Particularly, our results for the per-cell optimized $\alpha$ duplex scheme show uplink rate improvement of 42\% (as compared to the $33\%$ in~\cite{Alammouri2015}) and downlink rate improvement up to 80\%  (as compared to the $28\%$ in~\cite{Alammouri2015}). The results further show that explicit uplink and downlink performance should be considered for efficient design of FD communication in cellular networks.

The rest of this paper is organized as follows, the system model and methodology of analysis are presented in Section~\ref{sec:SystemModel}. Then,  Section~\ref{sec:Optimization} details   the optimization process and introduce other scheme for comparison purposes.  Numerical  results  and discussion are presented in Section~\ref{sec:Results} before concluding the paper in Section~\ref{sec:Conclusion}.\\

\section{System model}\label{sec:SystemModel}
\subsection{Network Model}
\begin{figure}[t]
\begin{center}
\scalebox{0.32}{\includegraphics[]{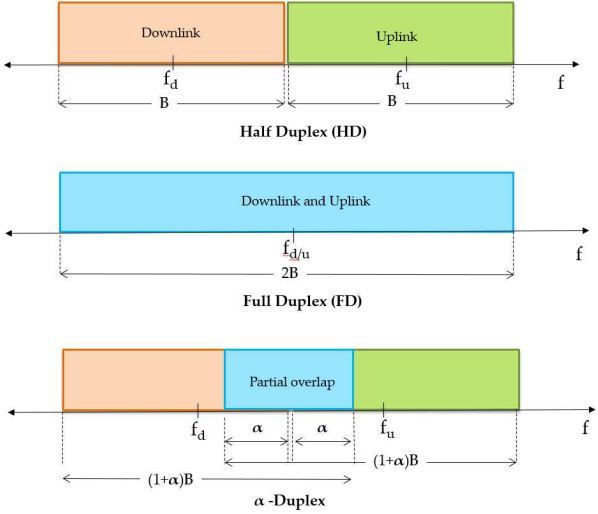}}
\end{center}
\caption{\small An illustrative diagram showing $\alpha$ duplex scheme.\normalsize}	
\label{fig:alphaDuplex_Sketch}
\end{figure}

\begin{figure*}[t]
\psfrag{Pui}[c][c][0.7]{~\! $p^{i}_{\text{u}}$}
\psfrag{Pdi}[c][c][0.7]{~\! $p^{i}_{\text{d}}$}
\psfrag{Puj}[c][c][0.7]{~\! $p^{j}_{\text{u}}$}
\psfrag{Pdj}[c][c][0.7]{~\! $p^{j}_{\text{d}}$}
\psfrag{cBSU(ii)}[c][c][0.6]{~\! $r_{\text{bu}}^{i,i},h_{\text{bu}}^{i,i},\alpha_{i}$}
\psfrag{cBSU(jj)}[c][c][0.6]{~\! $r_{\text{bu}}^{j,j},h_{\text{bu}}^{j,j},\alpha_{j}$}
\psfrag{cBSU(ji)}[c][c][0.6]{~\! $r_{\text{bu}}^{j,i},h_{\text{bu}}^{j,i}$}
\psfrag{cBSU(ij)}[c][c][0.6]{~\! $r_{\text{bu}}^{i,j},h_{\text{bu}}^{i,j}$}
\psfrag{cBSBS(ij)}[c][c][0.6]{~\! $r_{\text{bb}}^{i,j},h_{\text{bb}}^{i,j}$}
\psfrag{cUU(ij)}[c][c][0.6]{~\! $r_{\text{uu}}^{i,j},h_{\text{uu}}^{i,j}$}
\psfrag{ith cell}[c][c][0.7]{~\! $i^{th}$-cell}
\psfrag{jth cell}[c][c][0.7]{~\! $j^{th}$-cell}
\begin{center}
\scalebox{1.1}{\includegraphics[width=1\columnwidth]{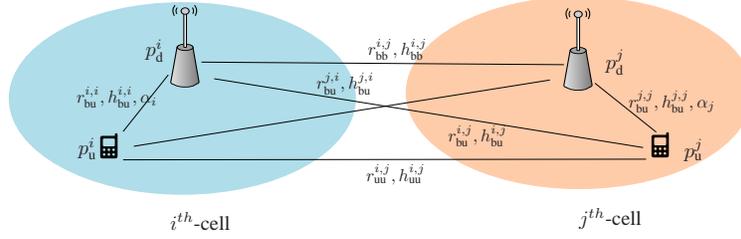}}
\end{center}
\caption{The system model representing two cells $i$ and $j$.}	
\label{fig:SystemModel}
\end{figure*}

We consider single-tier cellular network constituted by $N$ BSs. Each BS is serving one user who is uniformly distributed in its coverage range. All BSs and users are equipped with full-duplex transceivers with perfect self-interference cancellation (SIC) capability. It is worth mentioning that perfect SIC is assumed to focus on the effect of cross-mode interference on the network performance.  We assume one uplink and one downlink HD channels, each has $B$ Hz BW, that are universally reused across the network. Instead of assuming a common FD channel with $ 2 B$ Hz, we allow a flexible partial overlap between uplink and downlink channels.  That is, according to the parameter $\alpha \in [0,1]$, the uplink and downlink channels are expanded to $(1+\alpha) B$ such that they have an overlap of $2 \alpha B$. On one extreme, setting $\alpha = 0$ captures the HD scheme with $B$ Hz uplink channel, $B$ Hz downlink channel, and zero overlap. On the other extreme,  setting $\alpha = 1$ captures the FD scheme with a common (i.e., $2 B$ overlap) uplink and downlink channel with $2 B$ Hz. We also allow each BS to select its own overlap parameter $\alpha$ to be used in its cell. Hence, we define the overlap coefficient $\alpha_i$  which is used by the $i^{\text{th}}$ BS. An illustrative figure for the $\alpha$ duplex scheme is shown in Fig. \ref{fig:alphaDuplex_Sketch}. 

\subsection{Pulse-shaping \& Effective Cross-mode Interference}
As in \cite{Alammouri2015}, the uplink and downlink channels do not show rectangular shapes in the frequency domain\footnote{Rectangular pulses in the frequency domain are not used to avoid inter symbol interference in the time domain and to reduce the cross-mode interference with $\alpha$.}. Note that the pulse shape shown  in Fig. \ref{fig:alphaDuplex_Sketch} is just for illustration purpose. Instead, all uplink transmissions are assumed to use a unit energy Sinc$^2$(.) pulse shape and that all the downlink channels are assumed to use a unit power Sinc(.) pulse shape. Hence, the respective uplink and downlink pulse shapes in the frequency domain are given by:

\small
\begin{align}\label{eq:PulseShapes1}
S_{\text{u}}(f)=\frac{\rm{Sinc}^2\left(\frac{2 f}{(1+\alpha)B}\right)}{\sqrt{\int\limits_{- \infty}^{\infty} \rm{Sinc}^4\left(\frac{2 f}{(1+\alpha)B}\right) df}},
\end{align}
\normalsize
\small
\begin{align}\label{eq:PulseShapes2}
S_{\text{d}}(f)=\frac{\rm{Sinc}\left(\frac{2 f}{(1+\alpha)B}\right)}{\sqrt{\int\limits_{- \infty}^{\infty} \rm{Sinc}^2\left(\frac{2 f}{(1+\alpha)B}\right) df}}.
\end{align}
\normalsize

The effect of pulse shaping and partial uplink and downlink overlap can be captured by the effective interference factor, which measures the amount of cross-mode interference leakage from the matched filters at the receivers. Hence, for the uplink receivers (i.e. at the BSs), the cross-mode interference factor is given by
\small
\begin{equation}  \label{fac1}
\mathcal{C}_{\text{u}} (\alpha)=\int\limits_{-(1+\alpha)B/2}^{(1+\alpha)B/2} S_{\text{u}}^{*}(f) S_{\text{d}}(f-{(1-\alpha)B})df ,
\end{equation}
\normalsize
where the downlink pulse is shifted by the amount of $(1-\alpha)B$ due to the  frequency mismatch as a consequence of  the partial overlap. As reflected by \eqref{fac1}, for HD case the uplink and downlink pulses are separated by $B$, and for FD case they are perfectly aligned between uplink and downlink. It is also worth mentioning that the integration limits are due to the low pass filter at the receiver which captures the energy within the BW of interest only.  Similarly, for the downlink receivers (i.e., at the users), the cross-mode interference factor is given by
\small
\begin{equation}  \label{fac2}
\mathcal{C}_{\text{b}} (\alpha)=\int\limits_{-(1+\alpha)B/2}^{(1+\alpha)B/2} S_{\text{d}}^{*}(f) S_{\text{u}}(f-{(1-\alpha)B})df ,
\end{equation}
\normalsize
The cross-mode interference factors given in \eqref{fac1}, and \eqref{fac2} show the tradeoff imposed by the parameter $\alpha$ on the system performance. From one side, both $\mathcal{C}_{\text{u}}(\alpha)$ and $\mathcal{C}_{\text{b}}(\alpha)$ are increasing functions of $\alpha$ due to the larger overlap between the pulse shapes and the higher integration range, which implies higher cross-mode interference. On the other hand, looking into \eqref{eq:PulseShapes1} and \eqref{eq:PulseShapes2} we see that the effective BW accessed by uplink and downlink transmission also increases in $\alpha$, which implies higher rate. Hence, the overlap parameter $\alpha$ should be carefully tuned to balance the tradeoff between the channel BW and the cross-mode interference. As will be shown later, the uplink cross-mode interference factor $\mathcal{C}_{\text{u}} (\alpha)$ has a prominent effect on the uplink signal-to-interference-plus-noise-ratio (SINR) due to the high transmission powers of the BSs. In contrast,  $\mathcal{C}_{\text{b}} (\alpha)$  has minor effect of the downlink SINR due to the low transmission powers by the users.

Both $\mathcal{C}_{\text{u}}(\alpha)$ and $\mathcal{C}_{\text{b}}(\alpha)$ depend highly on the pulse shapes. $\rm{Sinc}$ and $\rm{Sinc}^2$  pulse shapes are respectively assumed for downlink and uplink all along this paper, as mentioned above; the optimization of those pulse shape is left for future  further analysis.

\subsection{SINR Representation}

To facilitate the SINR representation of the considered $\alpha$-duplex system, the distances $(r)$ and the channel power gains $(h)$  between any two transceivers in the network are arranged in a matrix format. The subscript indicates the type of the two transceivers ("b" for base station and "u" for user), and the superscript specifies their indices. $l(d)$ is  the attenuation due to distance (pathloss) for a distance  $d$, this function depends on the deployment scenario as shown in \cite{3gpp}.  $p_{\text{u}}^{i}$ ($p_{\text{b}}^{i}$) is the transmit power of the $i^{\text{th}}$ user (BS) and $\boldsymbol{p_{\text{u}}}$ ($\boldsymbol{p_{\text{b}}}$) is a vector containing the transmit powers of $N$ users (BSs). Fig.~\ref{fig:SystemModel} shows an example of the considered network with two cells $i$ and $j$.

The $i^{\text{th}}$  uplink signal-to-interference-plus-noise (SINR) is defined as
\footnotesize
\begin{align}
\gamma^{i}_{\text{u}}(\boldsymbol{p_{\text{u}}}, \boldsymbol{p_{\text{d}}},\boldsymbol{\alpha})&=\frac{p_{\text{u}}^{i}h_{\text{bu}}^{i,i}l(r_{\text{bu}}^{i,i})}{\displaystyle\sum_{j=1,j\neq i}^{N}p_{\text{b}}^{j}h_{\text{bb}}^{j,i}l(r_{\text{bb}}^{j,i})\vert C_{\text{u}}(\alpha_{j})\vert ^{2} +\!\!\displaystyle\sum_{j=1,j\neq i}^{N}p_{\text{u}}^{j}h_{\text{bu}}^{j,i}l(r_{\text{bu}}^{j,i})+\sigma_{\text{b}}^{i} (\alpha)},
\end{align}
\normalsize
where $\boldsymbol{\alpha}=\{\alpha_1, \alpha_2, ....,\alpha_N \}$ is the vector of all fractions of overlapping spectrums, and $\sigma_{\text{b}}^{i}(\alpha)$ is the noise variance at the $i^{\text{th}}$ BS receiver, explicitly $\sigma_{\text{b}}^{i}(\alpha)= (1+\alpha)BN_{o}$ where $N_{o}$ is the noise spectral density at the BS receiver . The first term in the denominator is the cross-mode interference. In this case, this is the interference from downlink transmission affecting the uplink transmission due to the amount of overlap in bandwidth.

In the  same fashion, the  $i^{\text{th}}$  downlink  SINR is:
\footnotesize
\begin{align}
\gamma^{i}_{\text{b}}(\boldsymbol{p_{\text{u}}}, \boldsymbol{p_{\text{d}}},\boldsymbol{\alpha})&=\frac{p_{\text{b}}^{i}h_{\text{bu}}^{i,i}l(r_{\text{bu}}^{i,i})}{\displaystyle\sum_{j=1,j\neq i}^{N}p_{\text{u}}^{j}h_{\text{uu}}^{j,i}l(r_{\text{uu}}^{j,i})\vert C_{\text{b}}(\alpha_{j})\vert ^{2} +\displaystyle\sum_{j=1,j\neq i}^{N}p_{\text{b}}^{j}h_{\text{bu}}^{j,i}l(r_{\text{bu}}^{j,i})+\sigma_{\text{u}}^{i} (\alpha)}.
\end{align}
\normalsize

Therefore, the uplink rate and downlink rate  for the  $i^{\text{th}}$ user are respectively
\small
\begin{align}
R^{i}_{\text{u}}(\boldsymbol{p_{\text{u}}}, \boldsymbol{p_{\text{d}}},\boldsymbol{\alpha} )&=(1+\alpha_{i})B\log(1+\gamma^{i}_{\text{u}}(\boldsymbol{\alpha})), \text{ and}\\
R^{i}_{\text{b}}(\boldsymbol{p_{\text{u}}}, \boldsymbol{p_{\text{d}}},\boldsymbol{\alpha} )&=(1+\alpha_{i})B\log(1+\gamma^{i}_{\text{b}}(\boldsymbol{\alpha})).
\end{align}
\normalsize

%


\section{Optimization Problem}\label{sec:Optimization}
\subsection{The Suggested Scheme}
In this paper, we consider maximizing an overall network utility to determine the uplink/downlink transmission powers and the spectrum overlap fraction between the uplink and downlink spectrum in each cell. The focus is on two types of objective functions that are used in analysis and performance illustration: sum-rate and sum log-rate. For instance, the sum-rate problem can be formulated as follows:

\small
\begin{subequations}\label{eq:Optimization_Sum}
\begin{align}
 \underset{\boldsymbol{p_{\text{u}}}, \boldsymbol{p_{\text{d}}},\boldsymbol{\alpha} }{\tr{max}} \quad &\sum_{i=1}^{N}R^{i}_{\text{b}}(\boldsymbol{p_{\text{u}}}, \boldsymbol{p_{\text{d}}},\boldsymbol{\alpha})+R^{i}_{{\text{b}}}(\boldsymbol{p_{\text{u}}}, \boldsymbol{p_{\text{d}}},\boldsymbol{\alpha}),\label{eq:ObjectiveSum}\\
\tr{s.t.\ \ } 			&\textbf{0}\leq \boldsymbol{p_{\text{u}}} \leq \boldsymbol{p}^{\text{max}}_{\text{u}},\\
& p_{\text{b}} \geq \boldsymbol{p}^{\text{min}}_{\text{b}},\label{eq:PminBS}\\
			&  \sum_{i=1}^{N}p^{i}_{\text{b}} \leq p^{\text{tot}}_{\text{b}},\label{eq:PtotBS}\\
			& \boldsymbol{\alpha}_{\text{min}}\leq \boldsymbol{\alpha} \leq \textbf{1},
\end{align}
\end{subequations}
\normalsize
where the optimization is over the powers $p_{\text{u}}$, $p_{\text{b}}$, and $\boldsymbol{\alpha}$, where $p^{\text{max}}_{\text{u}}$ is the maximum transmit power for users. Further, knowing that the users receiver is less sophisticated than the BSs' receiver, we assume that the BS has a minimum transmit power that is used to guarantee the coverage ( i.e. SINR) of the user at the edge of every cell $p^{\text{min}}_{\text{b}}$  \eqref{eq:PminBS}. The constraint \eqref{eq:PtotBS} represents the total transmit power constraint across all BSs. Note that the maximum transmit power affordable by users equipment are small compared to the one for BSs ($Np^{\text{max}}_{\text{u}}\ll p^{\text{tot}}_{\text{b}}$).   Based on the pulse shapes discussed earlier, it shown in \cite{Alammouri2015} that at $\alpha=0.275$, orthogonality is achieved  between the downlink and uplink such that $\mathcal{C}_{\text{u}}(0.275)=0$. However, $\mathcal{C}_{\text{u}}(0.275)>0$.  Due to the vulnerability of uplink transmission to downlink interference, and the negligibility of the interference from uplink to downlink., we take $\alpha=0.275$ as minimum value of  of the spectrum overlap.

The value of  $\mathcal{C}_{\text{u}}(\alpha)$ and $\mathcal{C}_{\text{b}}(\alpha)$ are approximated by polynomials  for $\alpha>0.275$.

\subsection{Proposed Solution}
The above optimization problem is a non-convex problem due to the  coupled interference terms in the SINR expressions. This paper, however, applies the interior point method to solve it \cite{Boyd2004}. While the suggested method does not guarantee global optimality because of the non-convexity of the original problem, the simulations section shows that the utilized method outperforms the benchmark schemes presented in the subsection \ref{subsec:Benchmarking}.

The following is a brief description of the utilized method. All details can be found in \cite{Boyd2004}. First, let $f_{i}(\boldsymbol{p_{\text{u}}}, \boldsymbol{p_{\text{d}}},\boldsymbol{\alpha} )\leq 0$ be the $i ^{\text{th}}$ scalar inequality constraint of the above optimization problem (There are $m\!\!=\!\!5N+1$ such constraints). Our optimization problem can be written as the following minimization problem

\small
\begin{subequations}
\begin{align}\label{eq:OptimizationSimpleForm}
 \underset{\boldsymbol{p_{\text{u}}}, \boldsymbol{p_{\text{d}}},\boldsymbol{\alpha} }{\tr{min}} & - \bigg[ \quad \sum_{i=1}^{N}R^{i}_{\text{b}}(\boldsymbol{p_{\text{u}}}, \boldsymbol{p_{\text{d}}},\boldsymbol{\alpha} )+R^{i}_{\text{u}}(\boldsymbol{p_{\text{u}}}, \boldsymbol{p_{\text{d}}},\boldsymbol{\alpha} )\bigg]\\
\tr{s.t.\ \ } 		&	f_{i}(\boldsymbol{p_{\text{u}}}, \boldsymbol{p_{\text{d}}},\boldsymbol{\alpha} ) \leq 0, \ \  i=1,...m.
\end{align}
\end{subequations}
\normalsize
Or equivalently, using the logarithmic barrier function\small
\begin{align}\label{eq:OptimizationSimpleForm2}
 \underset{\boldsymbol{p_{\text{u}}}, \boldsymbol{p_{\text{d}}},\boldsymbol{\alpha} }{\tr{min}} \   -\Bigg(\!\sum_{i=1}^{N}R^{i}_{\text{b}}(\boldsymbol{p_{\text{u}}}, \boldsymbol{p_{\text{d}}},\boldsymbol{\alpha} )+R^{i}_{\text{u}}(\boldsymbol{p_{\text{u}}}, \boldsymbol{p_{\text{d}}},\boldsymbol{\alpha} )\nonumber\\
 +\frac{1}{\tau} \sum_{i=1}^{m} \!\log(-f_{i}(\boldsymbol{p_{\text{u}}}, \boldsymbol{p_{\text{d}}},\boldsymbol{\alpha} ))\Bigg),
\end{align}
\normalsize

The main idea of the barrier method is  to solve \eqref{eq:OptimizationSimpleForm2} for a fixed $\tau$ at every iteration. The value of $\tau$ increases  at every iteration, until $\frac{1}{\tau}$ becomes less than a certain tolerance value. In our case, within every iteration, we use Newton's method with line search (i.e, we determine how much to move in the obtained direction) to solve \eqref{eq:OptimizationSimpleForm2}. 
%

The other utility function considered, i.e., the log-sum rate, comes as an alternative solution to the sum-rate which often leads to unbalanced performance gains between the uplink and downlink. The log-sum rate, on the other hand, presents a fairness alternative that relatively balances the downlink and uplink performance, as shown later in the simulations section. In this case, instead of utilizing the sum-rate function \eqref{eq:ObjectiveSum} in the above optimization problem, we utilize the following sum log-rate function:

\small
\begin{equation}
\underset{\boldsymbol{p_{\text{u}}}, \boldsymbol{p_{\text{d}}},\boldsymbol{\alpha} }{\tr{max}} \quad \sum_{i=1}^{N}\log\left(R^{i}_{\text{b}}(\boldsymbol{p_{\text{u}}}, \boldsymbol{p_{\text{d}}},\boldsymbol{\alpha} \right) +\log\left(R^{i}_{\text{u}}(\boldsymbol{p_{\text{u}}}, \boldsymbol{p_{\text{d}}},\boldsymbol{\alpha} )\right). \label{eq:Objective_log}
\end{equation}
\normalsize
The interior point method steps used to solve the sum-log rate function are omitted as they mirror the steps used to solve the sum-rate problem.

 \subsection{Benchmark Schemes}\label{subsec:Benchmarking}
  To illustrate the performance of our proposed scheme, we present here three benchmark techniques. 
  \begin{itemize}
  \item First, the traditional HD with power control,  obtained by optimizing \eqref{eq:Optimization_Sum}  and \eqref{eq:Objective_log}  by fixing $\boldsymbol{\alpha}\!\!=\!\!\boldsymbol{0}$.
  \item Second, the FD scheme with power control, obtained by optimizing \eqref{eq:Optimization_Sum}  and \eqref{eq:Objective_log}  by fixing $\boldsymbol{\alpha}\!\!=\!\!\boldsymbol{1}$ .
  \item  For the third scheme, every couple  \{BS,user\} uses the same spectrum overlap coefficient $\alpha=0.275 $, as proposed in \cite{Alammouri2015}. Further, every entity transmits with a fixed transmit power. For BSs, use $\frac{p^{\text{tot}}_{\text{b}}}{N}$; for users, use $p^{\text{max}}_{\text{u}}$.
  \end{itemize}

\section{Numerical results}\label{sec:Results}
\subsection{Simulation Setup and Results}
\begin{figure}[h]
\psfrag{--------AD----------}[c][c][0.9]{~\!$\alpha$-D}
\psfrag{--------FD----------}[c][c][0.9]{~\! FD}
\psfrag{--------HD----------}[c][c][0.9]{~\! HD}
\psfrag{a in [r], fixed powers}[c][c][0.8]{~\! $\alpha$ from \cite{Alammouri2015},  fixed power}
\psfrag{--------Uplink------}[c][c][0.9]{~\! Uplink}
\psfrag{--------Downlink------}[c][c][0.9]{~\! Downlink}
\psfrag{(N*PmaxUser)/PtotBS}[c][c][1.4]{~\! $NP_{\text{U}}^{\text{max}}/ P_{\text{BS}}^{\text{tot}}$}
\psfrag{Bits/sec/Hz}[c][c][1.4]{~\! Total rate per user, per unit HD-BW}
\begin{center}
\scalebox{0.50}{\includegraphics{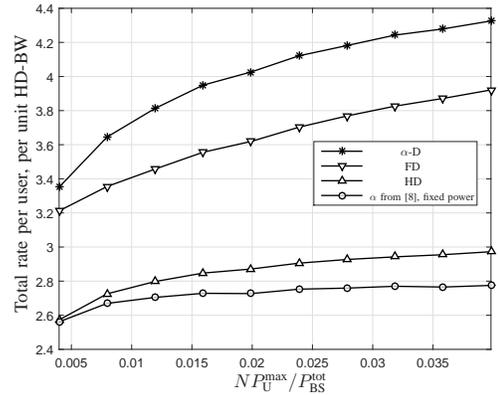}}
\caption{\small Variation of the average total rate ($R^{i}_{\text{u}}+R^{i}_{\text{d}}$) per user,  per unit HD-BW ($B$), with to the ratio of the uplink to downlink transmission power  for the sum rate maximization.}
\label{fig:SumRateSUM}
\end{center}
\end{figure}
\normalsize
\begin{figure}[h]
\psfrag{--------AD----------}[c][c][0.9]{~\!$\alpha$-D}
\psfrag{--------FD----------}[c][c][0.9]{~\! FD}
\psfrag{--------HD----------}[c][c][0.9]{~\! HD}
\psfrag{a in [r], fixed powers}[c][c][0.8]{~\! $\alpha$ from \cite{Alammouri2015},  fixed power}
\psfrag{--------Uplink------}[c][c][0.9]{~\! Uplink}
\psfrag{--------Downlink------}[c][c][0.9]{~\! Downlink}
\psfrag{(N*PmaxUser)/PtotBS}[c][c][1.4]{~\! $NP_{\text{U}}^{\text{max}}/ P_{\text{BS}}^{\text{tot}}$}
\psfrag{Bits/se/Hz}[c][c][1.4]{~\! Rate per user, per unit HD-BW}
\begin{center}
\scalebox{0.50}{\includegraphics{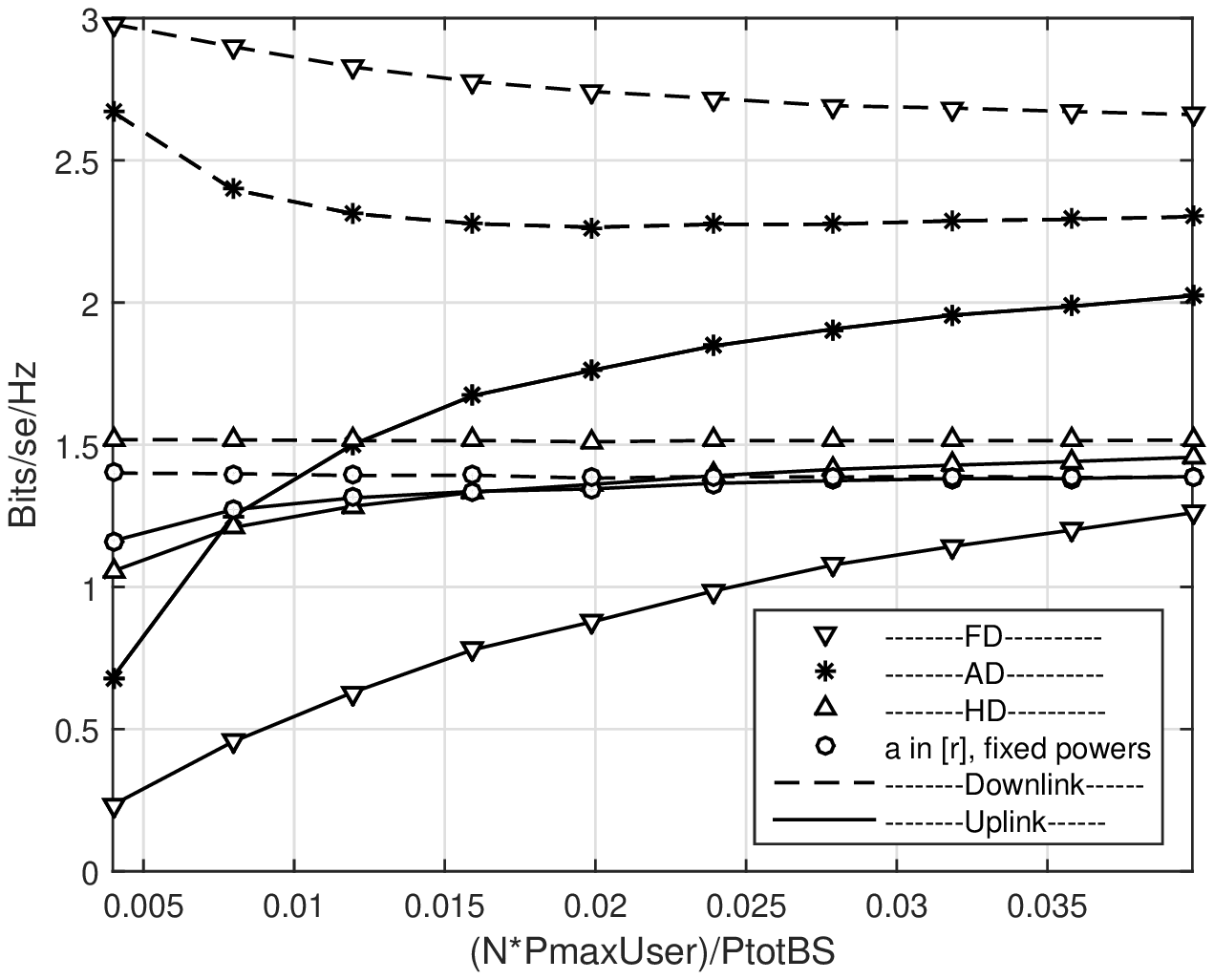}}
\caption{\small Variation of the rate per user, per unit HD-BW ($B$), with  the ratio of the uplink to downlink transmission power  for the sum rate maximization.}
\label{fig:SumRate}
\end{center}
\end{figure}
\normalsize
\begin{figure}[h]
\psfrag{--------AD----------}[c][c][0.9]{~\!$\alpha$-D}
\psfrag{--------FD----------}[c][c][0.9]{~\! FD}
\psfrag{--------HD----------}[c][c][0.9]{~\! HD}
\psfrag{a in [r], fixed powers}[c][c][0.8]{~\! $\alpha$ from \cite{Alammouri2015},  fixed power}
\psfrag{--------Uplink------}[c][c][0.9]{~\! Uplink}
\psfrag{--------Downlink------}[c][c][0.9]{~\! Downlink}
\psfrag{(N*PmaxUser)/PtotBS}[c][c][1.4]{~\! $NP_{\text{U}}^{\text{max}}/ P_{\text{BS}}^{\text{tot}}$}
\psfrag{Bits/se/Hz}[c][c][1.4]{~\!  Total rate per user, per unit HD-BW}
\begin{center}
\scalebox{0.50}{\includegraphics{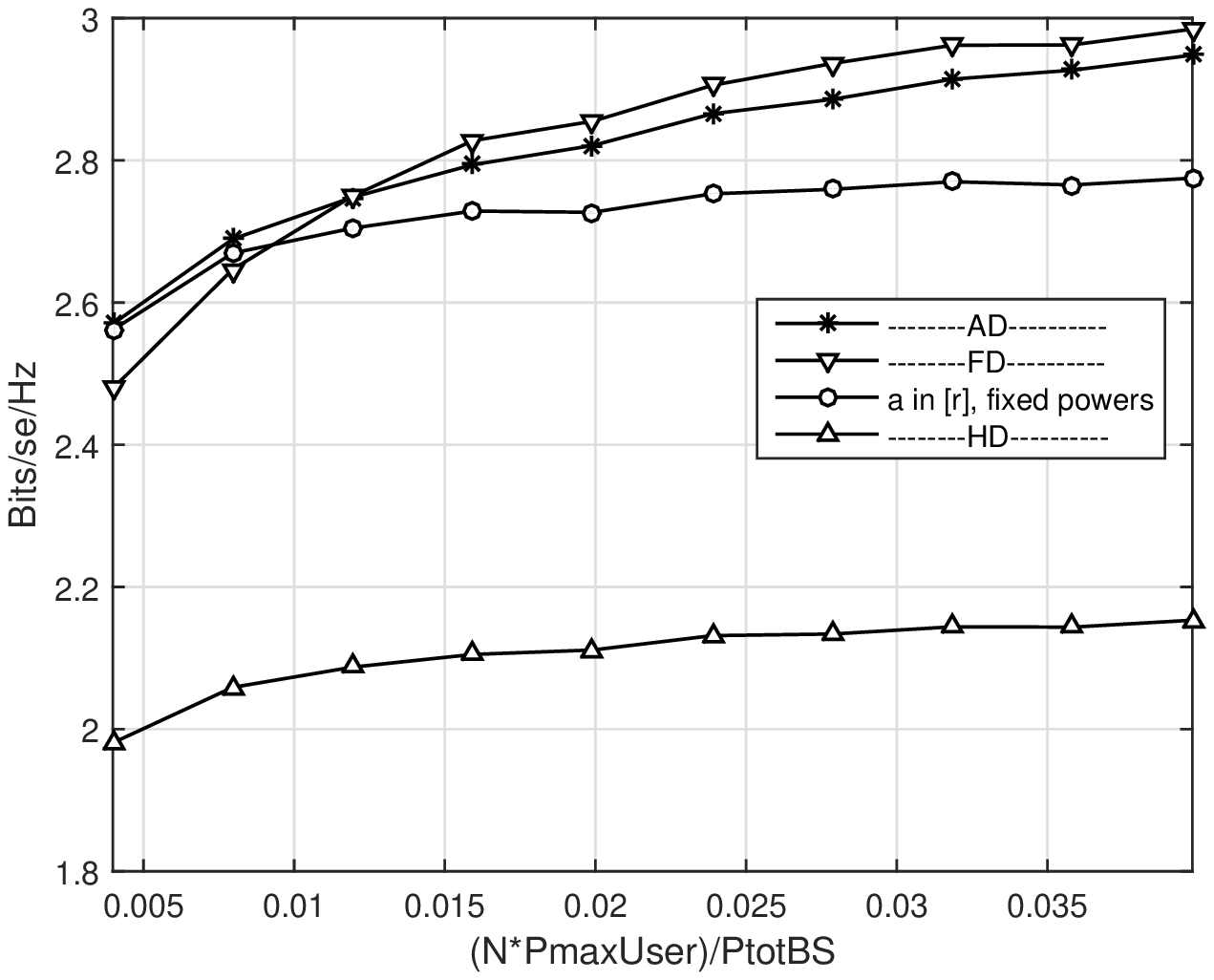}}
\caption{\small Variation of the average total rate ($R^{i}_{\text{u}}+R^{i}_{\text{d}}$) per user,  per unit HD-BW ($B$), with to the ratio of the uplink to downlink transmission power  for the sum of log  rate maximization.}
\label{fig:LogRateSUM}
\end{center}
\end{figure}
\normalsize
\begin{figure}[h]
\psfrag{--------AD----------}[c][c][0.9]{~\!$\alpha$-D}
\psfrag{--------FD----------}[c][c][0.9]{~\! FD}
\psfrag{--------HD----------}[c][c][0.9]{~\! HD}
\psfrag{a in [r], fixed powers}[c][c][0.8]{~\! $\alpha$ from \cite{Alammouri2015},  fixed power}
\psfrag{--------Uplink------}[c][c][0.9]{~\! Uplink}
\psfrag{--------Downlink------}[c][c][0.9]{~\! Downlink}
\psfrag{(N*PmaxUser)/PtotBS}[c][c][1.4]{~\! $NP_{\text{U}}^{\text{max}}/ P_{\text{BS}}^{\text{tot}}$}
\psfrag{Bits/se/Hz}[c][c][1.4]{~\!  Rate per user, per unit HD-BW}
\begin{center}
\scalebox{0.50}{\includegraphics{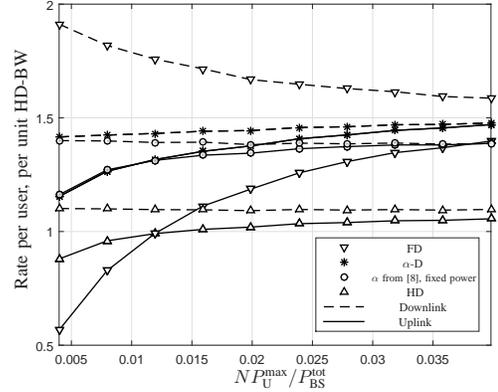}}
\caption{\small Variation of the rate per user, per unit HD-BW ($B$), with to the ratio of the uplink to downlink transmission power  for the sum of   log rate maximization.}
\label{fig:LogRate}
\end{center}
\end{figure}
\normalsize
In this section, the performance in terms of rate per user for FD, HD, and $\alpha$ duplex schemes are evaluated and compared. For the simulations, we consider an urban macro-cell environment in which the network and propagation model are employed according to 3GPP standard in \cite{3gpp}. Specifically, the used pathloss model is $l(d)=22 \log_{10}(d)+28+ 20\log_{10}(f_{c})$, where $d $ is the propagation distance in meters and $f_{c}$ is the carrier frequency in GHz. In each simulation run, $N\!\!=\!9$ BSs, with inter site distance of 500 m, are deployed. Then, a user is randomly dropped inside the coverage area of each BS. The channels BSs are selected to be 20 MHz on each direction.

Fig.~\ref{fig:SumRateSUM} shows the optimal rate for the sum rate maximization scheme in \eqref{eq:Optimization_Sum} with the ratio of the uplink to downlink transmission powers. The figure confirms the superiority of the $\alpha$ duplex scheme over all other schemes. Particularly, the $\alpha$ duplex scheme offers $43 \%$ sum rate gain over the traditional HD scheme and $10\%$ sum rate gain over the FD case. The figure also shows a $30\%$ sum rate gain of the FD scheme over the traditional HD scheme. The effect of the transmit power disparity between the uplink and downlink is also highlighted by  Fig.~\ref{fig:SumRateSUM}. The figure shows that increasing the maximum transmission power of  the uplink leads to a higher FD and $\alpha$-duplex performance, while the HD scheme saturates. The reason behind this behavior can be attributed to the vulnerability of the uplink and the prominent effect of the downlink to uplink interference. Increasing the uplink power provides more balanced operation between the uplink and downlink, which enables higher spectrum overlap and improves the overall performance.

Fig.~\ref{fig:SumRateSUM} shows an overall network performance, which may be misleading as it does not quantify the gain of the uplink and downlink separately. Therefore, we plot Fig.~\ref{fig:SumRate} to investigate the explicit uplink and downlink performances. The figure confirms the uplink vulnerability to downlink interference. When the uplink transmission power is low, $\alpha$ duplex gains are mainly in the downlink , while the uplink performance can be highly deteriorated, specially in the FD case. As the maximum transmission power of the uplink increases, more balanced operation between uplink and downlink is maintained and higher gains in the uplink is achieved. Note that the gains in the uplink come on a slight degradation in the downlink due to the cross mode interference. However, the overall performance increases as shown in  Fig.~\ref{fig:SumRateSUM}. The results in Fig.~\ref{fig:SumRate} also show the negligible effect of the uplink interference on the downlink performance. Hence, the FD scheme is the best for the downlink rate, while $\alpha$ duplex with optimized overlap between uplink and downlink balances the tradeoff between uplink and downlink performance and results in the highest overall network performance. Finally, Fig.~\ref{fig:SumRate} manifests the importance of explicitly accounting for the uplink and downlink performances, and shows that considering an overall network performance hides the  loss uplink may enconter.

Fig.~\ref{fig:SumRateSUM} shows that there is critical value of the uplink maximum transmission power (i.e., below 0.012 of the BS power) in which maximizing the sum rate through any uplink/downlink spectrum overlap scheme would always degrade the uplink performance, compared to the HD case. For instance, when the BS power is 200 times the maximum user terminal power,  $\alpha$ duplex scheme would deteriorate the uplink power with $25\%$ and FD scheme would deteriorate the uplink power with  $75\%$. This emphasizes the importance of fair rate utility function when there exists high disparity between uplink and downlink powers.

\begin{figure*}[t!]
    \centering
    \begin{subfigure}[t]{0.25\textwidth}
        \includegraphics[width=\textwidth]{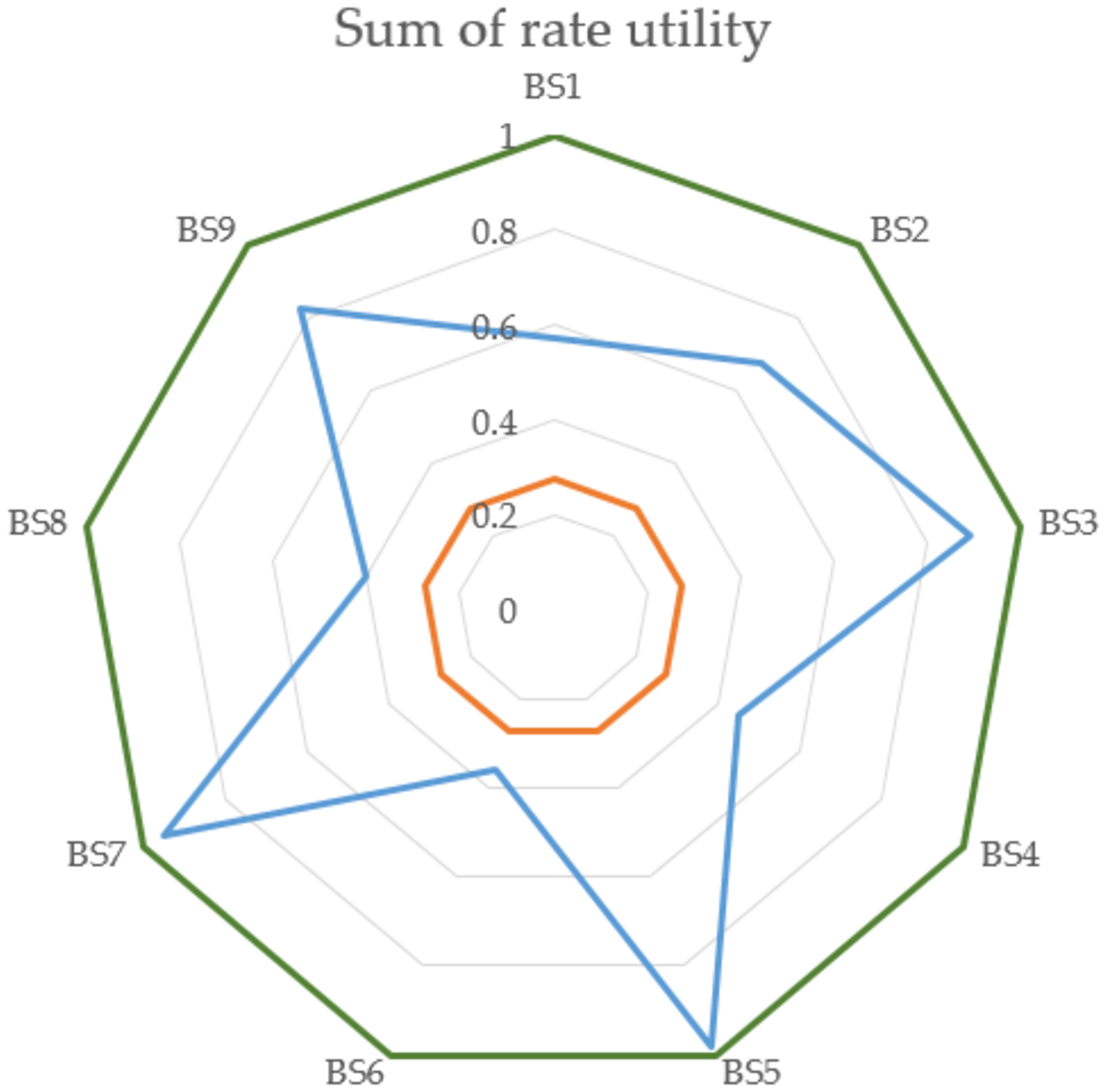}\vskip 9pt
                \includegraphics[width=\textwidth]{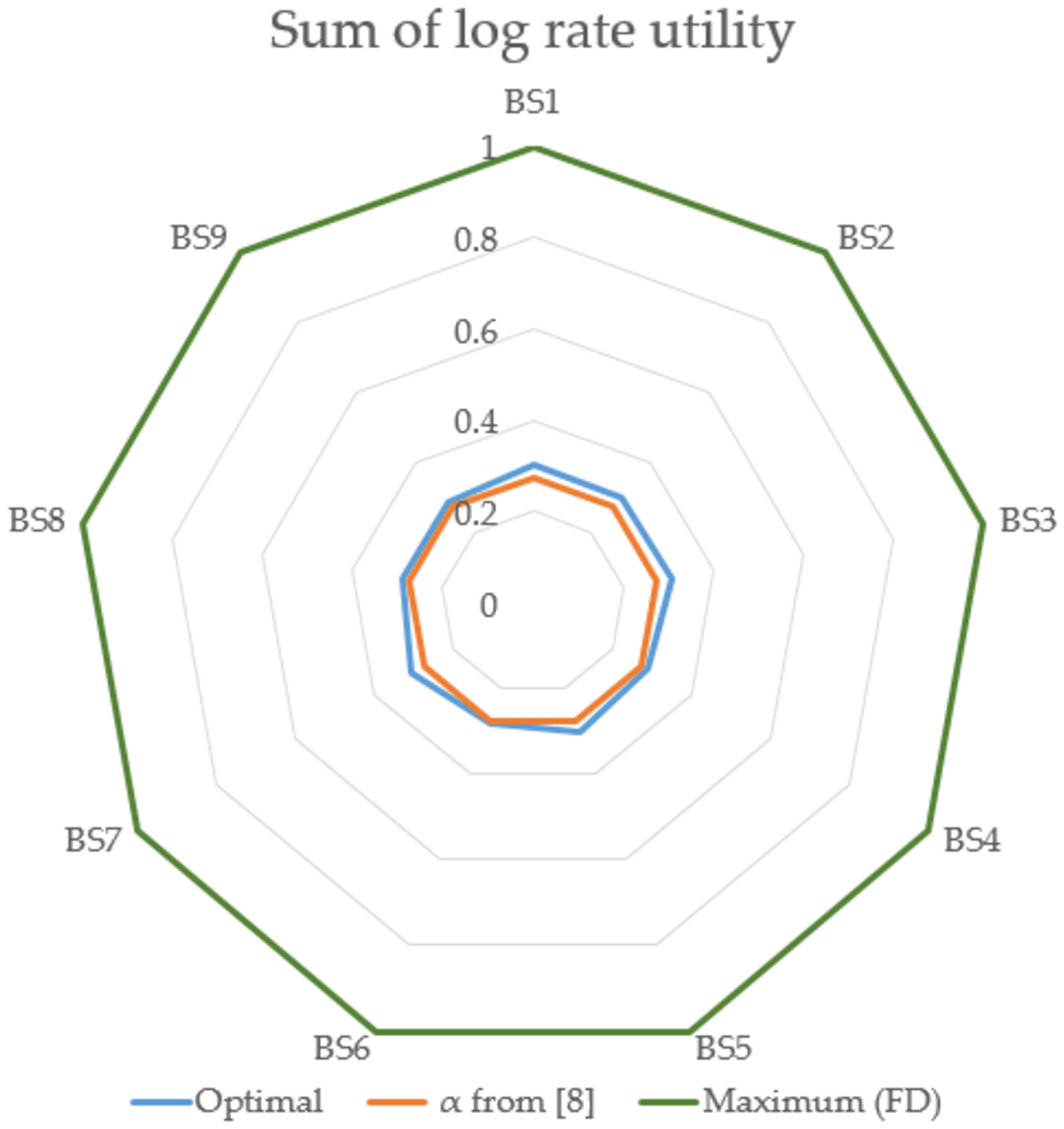}

        \caption{Low user transmit power\\\centering(100 mW)}
    \end{subfigure}\hspace{0.0625\textwidth}
    \begin{subfigure}[t]{0.25\textwidth}
        \includegraphics[width=\textwidth]{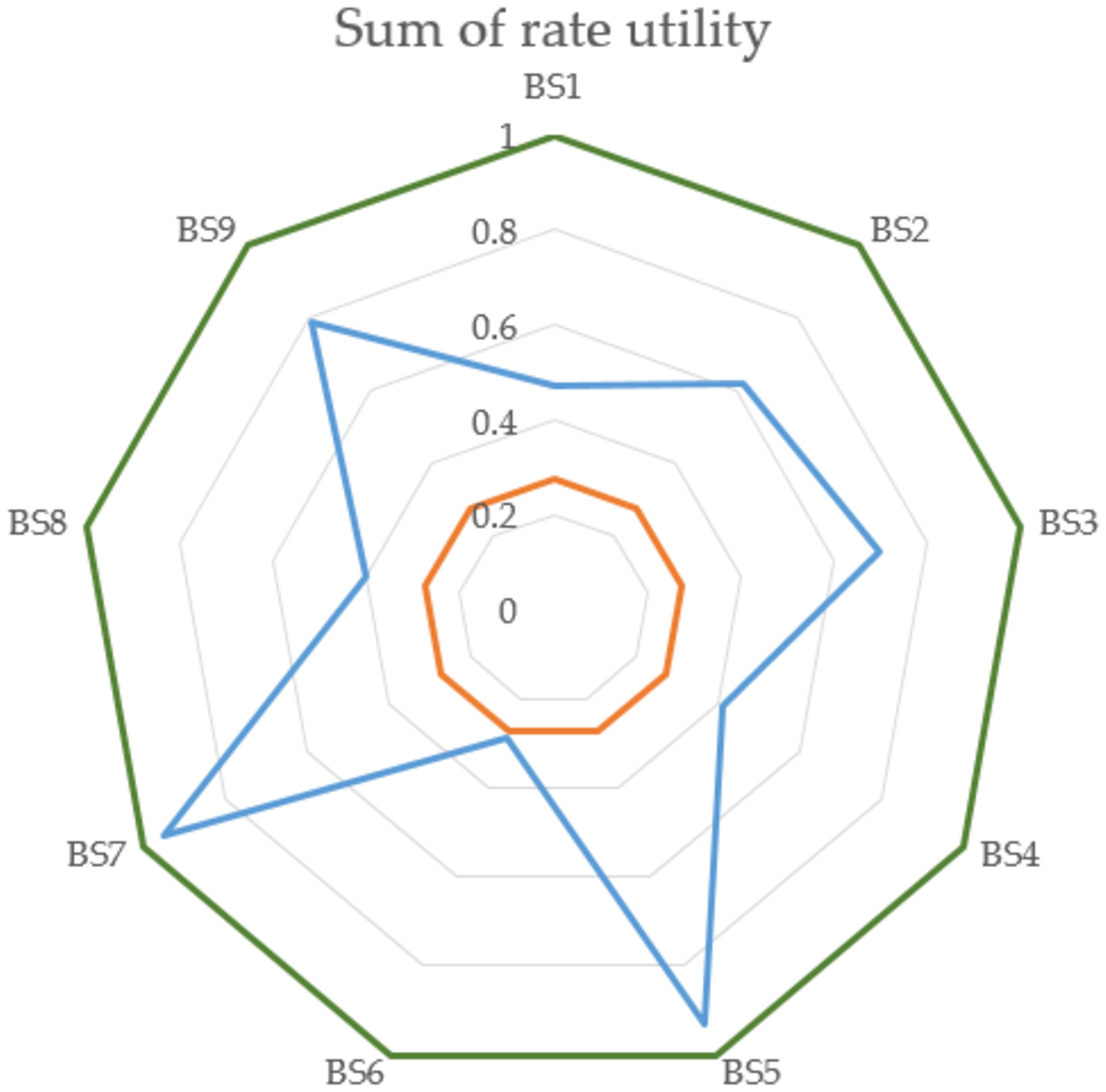}\vskip 9pt
                \includegraphics[width=\textwidth]{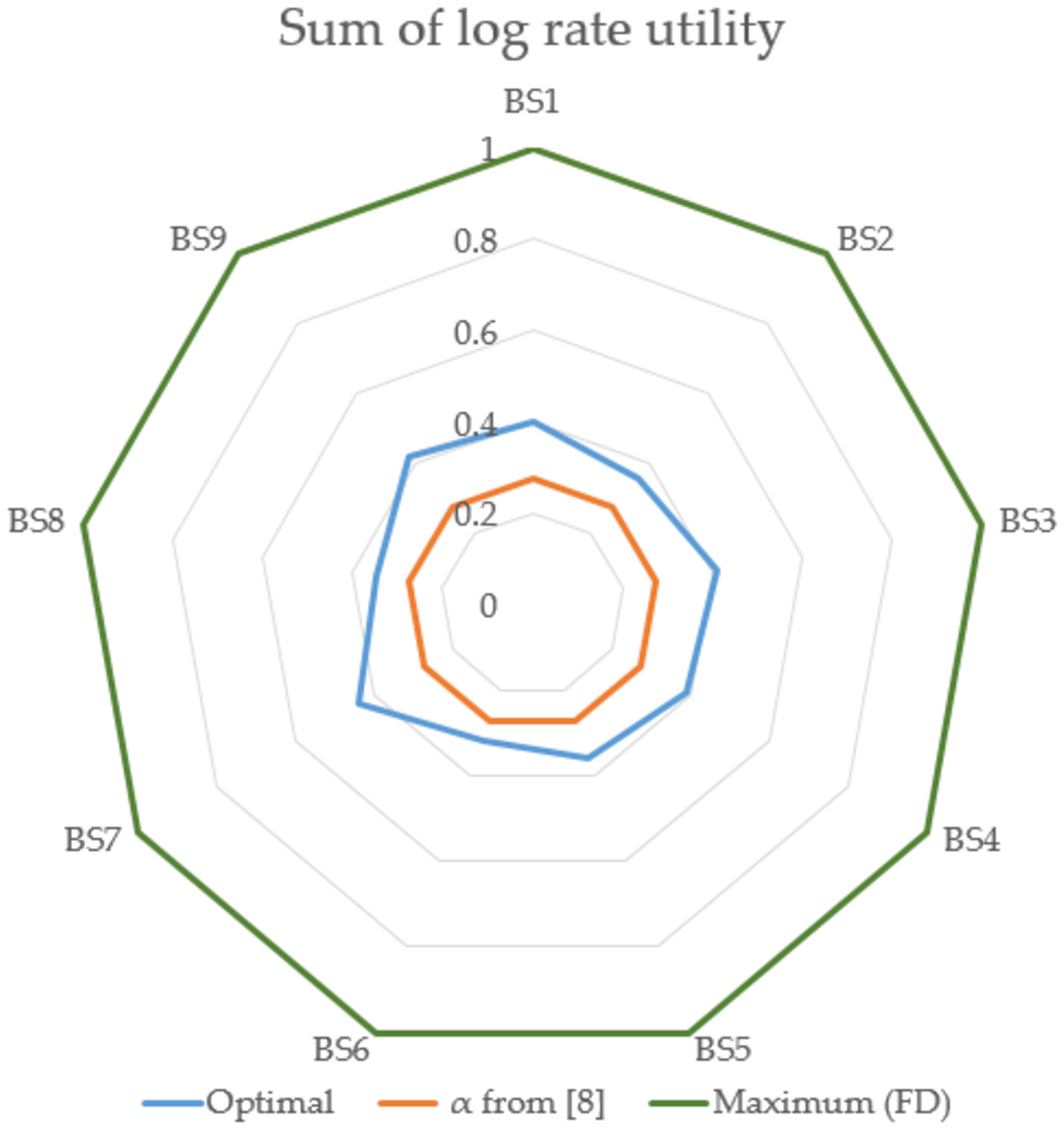}

             \caption{Medium user transmit power\\\centering(500 mW)}
    \end{subfigure}\hspace{0.0625\textwidth}
    \begin{subfigure}[t]{0.25\textwidth}
        \includegraphics[width=\textwidth]{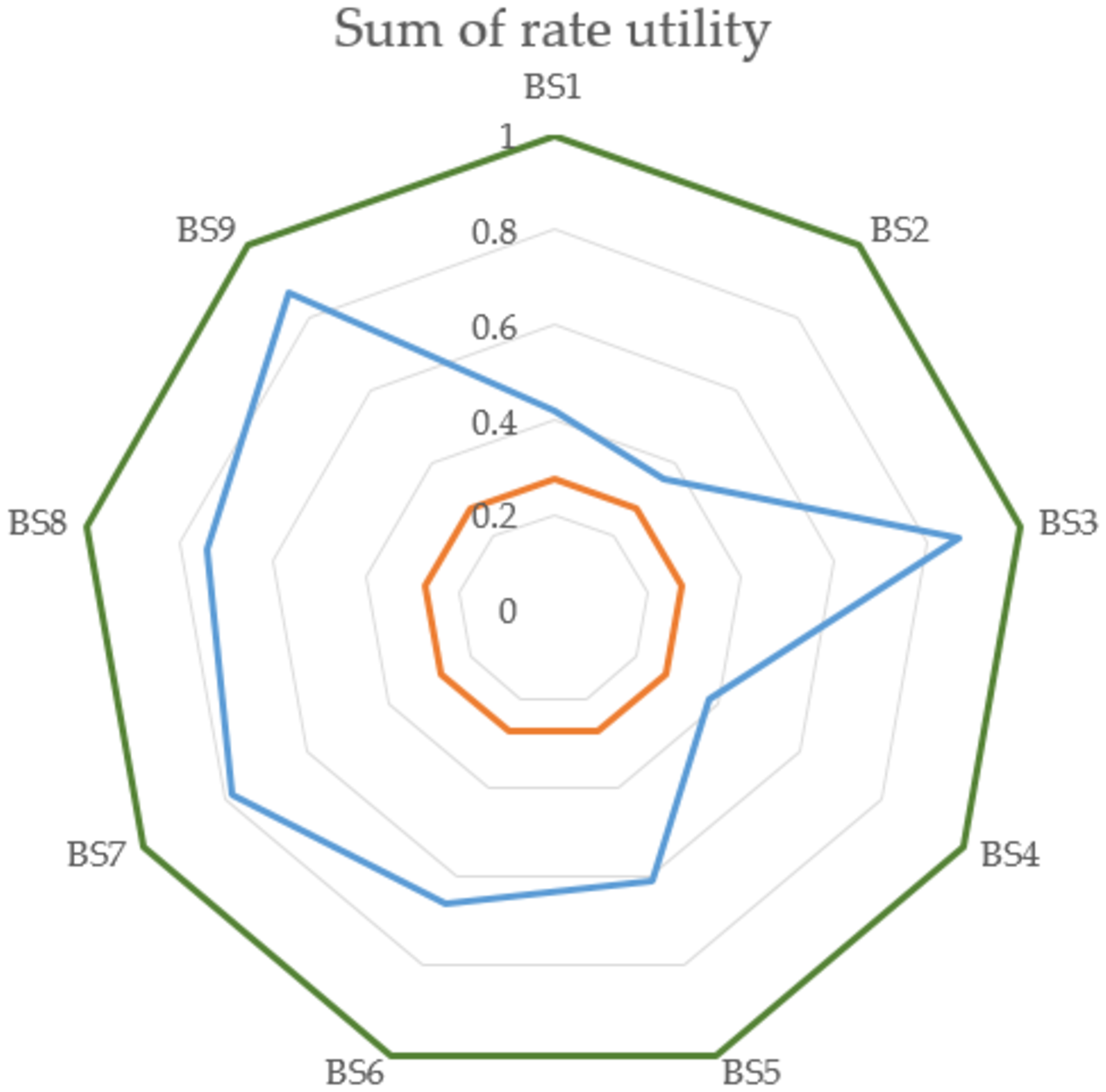}\vskip 9pt
                \includegraphics[width=\textwidth]{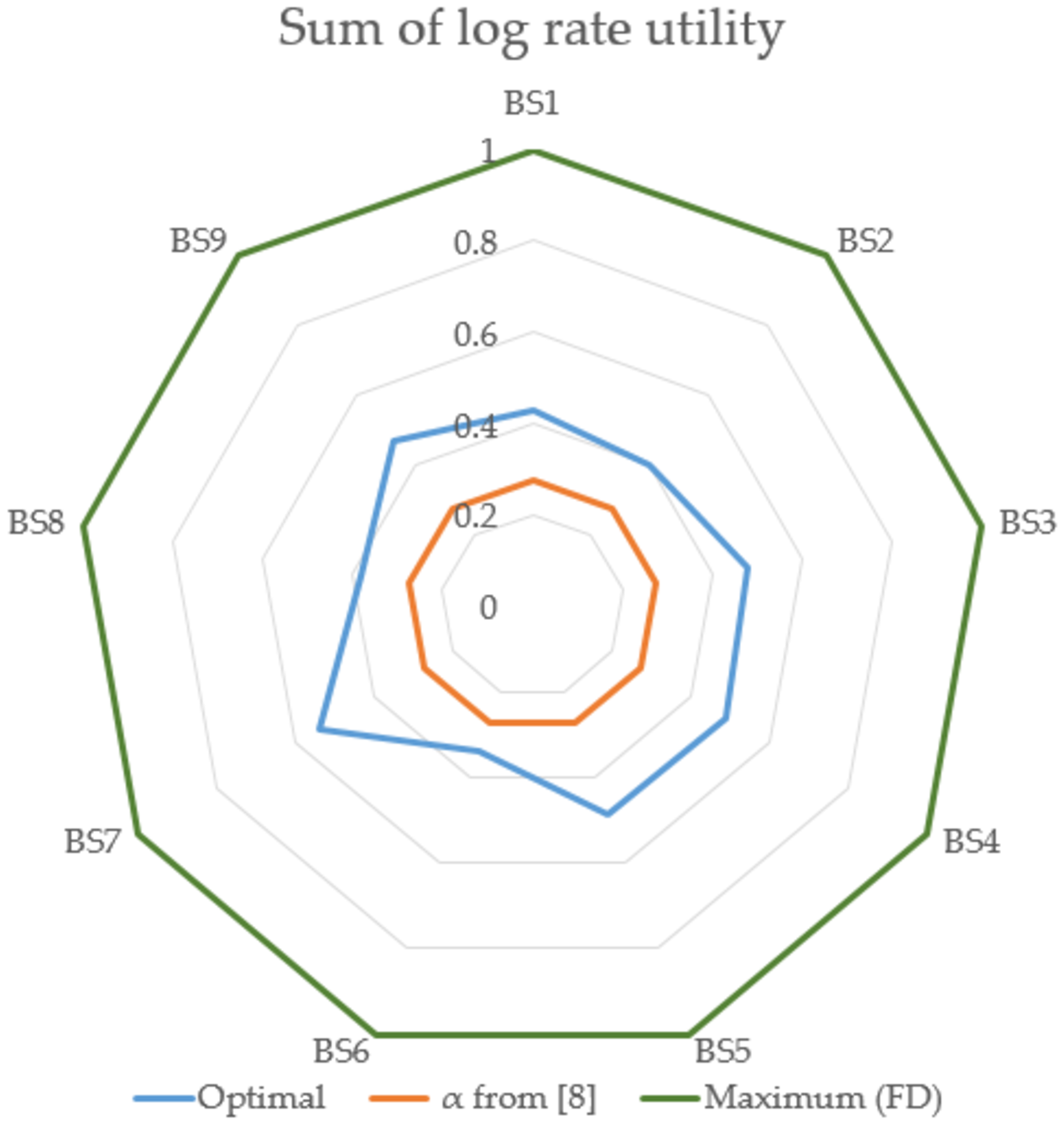}

                \caption{High user transmit power\\\centering(1 W)}
    \end{subfigure}
    \caption{Chart showing the optimal value of $\boldsymbol{\alpha}$ for different value of the maximum power at users}
    \label{fig:alphaselected}
\end{figure*}
Fig.~\ref{fig:LogRateSUM} and  \ref{fig:LogRate} show the rates obtained from solving the summation of logarithm of rates utility given in \eqref{eq:OptimizationSimpleForm2}. In terms of the overall network rate in Fig.~\ref{fig:LogRateSUM} , both FD and $\alpha$ duplex have comparable performance. However, for the explicit uplink and downlink performance in \ref{fig:LogRate}, the superiority of the $\alpha$ duplex in balancing the uplink and downlink performance is emphasized. The gain confirms the importance of accounting for the explicit uplink and downlink performance for system with overlapped spectrum access. The results in Fig.~\ref{fig:LogRate} also show that the FD operation may lead to performance deterioration for the uplink even for fair utility function. On the other hand, the $\alpha$-duplex scheme is able to maintain the uplink performance even for low values of the maximum uplink transmission power.

In order to gain further insight into the system operation and shed light on the fairness of the employed utility functions, we show Fig.~\ref{fig:alphaselected}. The figure visualizes the optimal $\boldsymbol{\alpha}$ selected by the BSs for each utility function at different disparity levels between the uplink and downlink transmission powers. The figure shows that in the case of sum rate utility function most of the BSs tend to select high values of $\alpha$ regardless of $p_{\text{U}}^{\text{max}}$, which explains the high deterioration in the uplink performance at high disparity levels of the uplink/downlink transmission powers (cf. Fig.~\ref{fig:SumRate}). On the other hand, the value of $\alpha$ selected by BSs in the sum of log rate utility function highly depends on the disparity level of the uplink/downlink transmit powers, which prevents the deterioration of the uplink rate. Therefore, increasing the uplink transmission powers for the sum log rate utility function enables higher optimal uplink/downlink overlap due to the increased reliability of the uplink transmissions. Last but not least, the figure confirms that a static overlap parameter for all BSs is not the optimal choice.

\subsection{Discussion}
The results confirm the vulnerability of the uplink performance to the downlink interference. Hence, FD operation with complete overlap between uplink and downlink channels results is beneficial to the downlink but ruinous for the uplink. Therefore the $\alpha$-duplex scheme is advocated for cellular networks, as well as for other wireless network with high disparity between the forward and reverse links transmit powers. The results confirm the superiority of the $\alpha$-duplex scheme in all case studies conducted in the paper for both the sum and explicit utilities.

In terms of the utility functions, when the maximum transmit powers of the users equipment is sufficiently high, sum rate utility formulation is recommended. In this case, the results show $30\%$ rate gain for optimizing sum rate when compared to optimizing the sum log rates.  On the other hand, when the maximum transmit powers of the users equipment is low, sum of log rates utility formulation is recommended. In this case, the results show that sum rate may result in significant uplink deterioration (e.g.,  $75\%$ for FD and $25\%$ for the $\alpha$ duplex), while positive gain is obtained for both of the uplink and downlink cases in the case of sum of the log rate for the $\alpha$ duplex scheme. Also, the deterioration for the FD case when maximum transmit powers of the users equipment is low decreases from $75\%$ for the sum rate to $14\%$ for the sum of log rate maximization.

Finally, it can be observed in all figures  that, choosing  a system wide constant value of $\alpha$ for all BSs, without  power control, can lead to waste of resources. 
For instance, in Fig. \ref{fig:SumRateSUM}, it performs worse than  HD in terms of global rate of the network. However, for low transmit power, a small advantage for uplink  is observable  (fig. \ref{fig:SumRate}), this is because the fixed value of alpha is chosen in a way to  minimize the downlink interference.

\section{Conclusion}\label{sec:Conclusion}
This paper considers a cellular wireless system with partial spectrum overlap between the downlink and uplink, as a means to devise smart spectrum sharing techniques between the downlink and uplink. The paper
 proposes an interference management scheme that controls per-cell transmission powers as well as the amount of overlap between uplink and downlink resources in order  to maximize  an overall network utility (sum-rate and sum log-rate). The problem is solved using interior point method. The results reaffirm that uplink performance may be significantly deteriorated in FD cellular networks due to the high disparity between base stations and user terminals powers. The proposed scheme shows 43\% performance gain when compared to the traditional HD scheme, and 10\% performance gain when compared to the FD scheme (both HD and FD are with power control).   Furthermore,  the sugested scheme is highlighted by 53\% gains when compared to static uplink/downlink overlap without interference management. 

We also conclude that the utility function formulation should   depend   on the disparity between the uplink and downlink transmission powers. For low uplink  transmission power, it is preferable to optimize the utility function harmonizing the uplink and downlink profits, in our case, the sum of the logarithm of rates. In contrast,  for high uplink transmission power, optimizing the summation of rate grants full exploitation of the resources, which leads to simultaneous uplink and downlink gains.

Last but not least, we show that accounting for an overall performance metric might be misleading, and we highlighted  the importance of  accounting for explicit uplink and downlink performance. For the future work, instead of assuming a static pulse shapes for all transmissions, optimal pulse shaping for interference management will be investigated.


%
%
%

\bibliographystyle{IEEEtran}
\nocite{*}
\bibliography{ICCReferences}


\end{document}